\documentclass[a4paper,10pt,conference]{IEEEtran}

\usepackage{amsmath}
\usepackage{amssymb}
\usepackage{newtxmath}
\usepackage{booktabs}
\usepackage{graphicx}
\usepackage{microtype}
\usepackage{multirow}
\usepackage{pgfplots}
\usepackage{tikz}
\usepackage{xcolor}
\usepackage{xspace}
\usepackage[sort]{cite}
\usepackage{url}
\usepackage{eso-pic}
\usetikzlibrary{arrows.meta,calc,positioning}
\usepgfplotslibrary{groupplots}
\pgfplotsset{compat=1.18}

\newcommand{\RQ}[1]{\textit{RQ}\textsubscript{#1}}
\newcommand{\RQAnswer}[2]{%
  \par\smallskip
  \noindent\fbox{\parbox{\dimexpr\linewidth-2\fboxsep-2\fboxrule\relax}{\textbf{Answer to \RQ{#1}.} #2}}
  \par\smallskip
}
\newcommand{\HeyRep}{Hey et al.\(_{\mathrm{rep}}\)\xspace}
\newcommand{\ours}{our approach\xspace}
\newcommand{\Ours}{Our approach\xspace}
\newcommand{\OurS}{Our Approach\xspace}
\newcommand{\cfg}[1]{\textit{#1}}
\newcommand{\TOP}{\operatorname{Top}}
\newcommand{\SetA}{\(\mathrm{Set}_A\)\xspace}
\newcommand{\SetB}{\(\mathrm{Set}_B\)\xspace}
\newcommand{\SetC}{\(\mathrm{Set}_C\)\xspace}
\newcommand{\SetBTextOnly}{\(\mathrm{Set}_B^{\text{text-only}}\)\xspace}
\newcommand{\SetCTextOnly}{\(\mathrm{Set}_C^{\text{text-only}}\)\xspace}
\newcommand{\SetValid}{\(\mathrm{Set}_{\mathrm{valid}}\)\xspace}
\newcommand{\SetTest}{\(\mathrm{Set}_{\mathrm{test}}\)\xspace}
\newcommand{\SetTestTextOnly}{\(\mathrm{Set}_{\mathrm{test}}^{\text{text-only}}\)\xspace}
\newcommand{\revision}[1]{{\footnotesize\texttt{#1}}}
\newcommand{\cmark}{$\checkmark$}

\title{Multimodal Requirements Traceability Link Recovery in Industrial Settings}

\author{%
\IEEEauthorblockN{Haoming Huang and Shinpei Hayashi}
\IEEEauthorblockA{Institute of Science Tokyo\\Tokyo, Japan\\\{haoming, hayashi\}@se.comp.isct.ac.jp}
\and
\IEEEauthorblockN{Kazuya Kohata, MyeongKi Kim, Yuki Kurihara}
\IEEEauthorblockA{YSL SOLUTION CO.,LTD.\\Kanagawa, Japan\\\{kazuya.kohata, kimu.myongi, yuki.kurihara\}@ysl.co.jp}
}
\begin{document}

\AddToShipoutPictureFG*{%
  \AtPageLowerLeft{%
    \put(\LenToUnit{\dimexpr1in+\oddsidemargin\relax},\LenToUnit{15mm}){%
      \begingroup
      \setlength{\fboxsep}{4pt}%
      \setlength{\fboxrule}{0.4pt}%
      \fbox{\parbox[b]{\dimexpr\textwidth-2\fboxsep-2\fboxrule\relax}{%
        \fontsize{7}{8}\selectfont
        \textcopyright\ 2026 IEEE. Personal use of this material is permitted. Permission from IEEE must be obtained for all other uses, in any current or future media, including reprinting/republishing this material for advertising or promotional purposes, creating new collective works, for resale or redistribution to servers or lists, or reuse of any copyrighted component of this work in other works.
      }}%
      \endgroup
    }%
  }%
}

\maketitle
\pagestyle{plain}
\thispagestyle{plain}
\addtolength{\footskip}{14mm}

\begin{abstract}\boldmath
Requirements traceability links support activities such as change-impact analysis, coverage checking, and validation.
Industrial requirements can include traceable images, which text-only approaches cannot process directly.
For a fixed classifier configuration, binary decisions alone offer no direct way to adjust the balance between missed and incorrect links.
We extend a retrieval-augmented generation (RAG) approach for inter-requirements traceability link recovery (TLR).
Our extension retrieves text and image candidates separately, classifies each retrieved pair with a vision-language model, and filters the predicted links by verbalized confidence.
We evaluate \ours on three non-overlapping industrial requirement sets and six public text-only datasets.
On the industrial test sets, \ours achieves an \(F_1\) score of 0.310, compared with 0.114 for our replication of the baseline approach by Hey et al.
\Ours recovers links in all four modality combinations while improving text-only \(F_1\) on the industrial data and maintaining similar performance on the public datasets, but recovery remains limited by failures to retrieve related text requirements for image queries.
Confidence filtering produces link sets with different precision and recall, removing false positives and some correctly recovered links from the unfiltered link set.
This study provides industrial evidence on the potential and limitations of a current RAG-based TLR approach for multimodal requirements and identifies cross-modal retrieval as the component to prioritize in further optimization.
\end{abstract}


\section{Introduction}\label{sec:introduction}

Requirements traceability records relations among requirements and other software artifacts.
In the inter-requirements traceability link recovery (TLR) setting of Hey et al.~\cite{HeyEtAl2025RAGRequirementsTLR}, links between high-level requirements (HLRs) and low-level requirements (LLRs) represent refinement or predecessor--successor relations.
A survey of requirements traceability~\cite{WinklerPilgrim2010TraceSurvey} reports that trace links can support activities such as change-impact analysis, coverage checking, validation, testing, and navigation.
An empirical study~\cite{MaderEgyed2015Benefits} shows that accurate trace links can improve both completion time and correctness in selected software-maintenance tasks.

Traceability adoption faces time, cost, and organizational barriers~\cite{RuizEtAl2023WhyTrace}.
Automated TLR must recover correct links while limiting the false-positive candidates that engineers need to review~\cite{HayesEtAl2006CandidateLinks,RuizEtAl2023WhyTrace}.

Existing retrieval-augmented generation (RAG) approaches to TLR include LiSSA~\cite{FuchssEtAl2025LiSSA} and Hey et al.'s inter-requirements approach~\cite{HeyEtAl2025RAGRequirementsTLR}.
They retrieve candidates and use a large language model (LLM) to classify each pair.
Existing evaluations of these RAG-based approaches use textual artifacts or artifacts converted to text before retrieval and classification.
Section~\ref{sec:related-work} reviews their designs and evaluation settings.

We focus on two practical challenges in industrial TLR\@.
The first challenge for these text-based approaches is handling images as traceable requirement elements.
An interview study of requirements practice~\cite{FranchEtAl2023RequirementsPractice} reports that natural language appeared in 23 of 24 studied industrial projects, while 20 practitioners used more than one specification artifact and 12 used graphical elements.
The reported graphical elements included mockups, diagrams, tables, and pictures.
Industrial requirements can include images that are themselves traceable elements.
When image elements are omitted from the input, links involving those elements cannot be recovered.
Engineers who need links for the full set of requirements must then check those elements separately.

\begin{figure}[b]\centering
  \resizebox{\columnwidth}{!}{\definecolor{figdark}{HTML}{17202A}
\definecolor{figtext}{HTML}{334E68}
\definecolor{figline}{HTML}{8A9AA8}
\definecolor{figgreen}{HTML}{1D7C4C}
\begin{tikzpicture}[
  font=\sffamily,
  card/.style={draw=figline, fill=white, rounded corners=1.5mm, line width=0.7pt},
  idbox/.style={draw=figline, fill=black!4, rounded corners=0.8mm, line width=0.5pt, font=\bfseries\scriptsize, text=figtext, inner xsep=3pt, inner ysep=1.5pt},
  label/.style={font=\bfseries\scriptsize, text=figtext, inner sep=0pt},
  body/.style={font=\small, text=figdark, align=left, inner sep=0pt},
  trace/.style={draw=figgreen, line width=1.2pt, -{Latex[length=2.4mm]}},
  trace label/.style={font=\bfseries\scriptsize, text=figgreen, fill=white, inner sep=1pt}
]
  \draw[card] (-1.55,-1.02) rectangle (1.55,1.02);
  \node[label, font=\bfseries\tiny, anchor=north west] at (-1.34,0.83) {HIGH-LEVEL REQUIREMENT};
  \node[idbox, anchor=north west] at (-1.34,0.48) {HA-7};
  \node[body, text width=2.55cm, anchor=north west] at (-1.34,0.04) {Set home address};

  \begin{scope}[xshift=0.9cm]
  \draw[card] (2.70,0.34) rectangle (7.05,2.05);
  \node[label, anchor=north west] at (2.91,1.86) {TEXT ELEMENT};
  \node[idbox, anchor=north west] (la3) at (2.91,1.51) {LA-3};
  \node[body, anchor=west] at ([xshift=5pt]la3.east) {Detailed-address field};

  \draw[card] (2.70,-2.18) rectangle (7.05,-0.30);
  \node[label, anchor=north west] at (2.91,-0.49) {IMAGE ELEMENT};
  \node[idbox, anchor=north west] (la12) at (2.91,-0.84) {LA-12};
  \node[body, anchor=west] at ([xshift=5pt]la12.east) {Address-entry screen};
  \draw[draw=figline, fill=black!2, line width=0.55pt] (3.19,-1.30) rectangle (6.72,-2.02);
  \draw[draw=figline, fill=black!10, line width=0.45pt] (3.19,-1.30) rectangle (6.72,-1.49);
  \node[font=\bfseries\tiny, anchor=west, text=figtext] at (3.31,-1.395) {Set address};
  \node[font=\tiny, anchor=west, text=figtext] at (3.31,-1.65) {Street address};
  \draw[draw=figline, fill=white, line width=0.4pt] (4.70,-1.55) rectangle (6.47,-1.73);
  \node[font=\tiny, anchor=west, text=figtext] at (4.74,-1.64) {1-1-1 Chiyoda};
  \node[font=\tiny, anchor=west, text=figtext] at (3.31,-1.90) {Postal code};
  \draw[draw=figline, fill=white, line width=0.4pt] (4.70,-1.80) rectangle (5.70,-1.98);
  \node[font=\tiny, anchor=west, text=figtext] at (4.74,-1.89) {100-0001};
  \draw[draw=figline, fill=black!7, rounded corners=0.4mm, line width=0.4pt] (5.80,-1.80) rectangle (6.59,-1.98);
  \node[font=\bfseries\tiny, text=figtext] at (6.195,-1.89) {Confirm};
  \end{scope}

  \draw[trace] (1.55,0.35) -- node[trace label, above, sloped] {refinement} (3.60,1.18);
  \draw[trace] (1.55,-0.35) -- node[trace label, below, sloped] {refinement} (3.60,-1.22);
\end{tikzpicture}}
  \caption{Illustrative inter-requirements trace links from one high-level requirement to text and image elements in the low-level requirements.}\label{fig:tlr-example}
\end{figure}
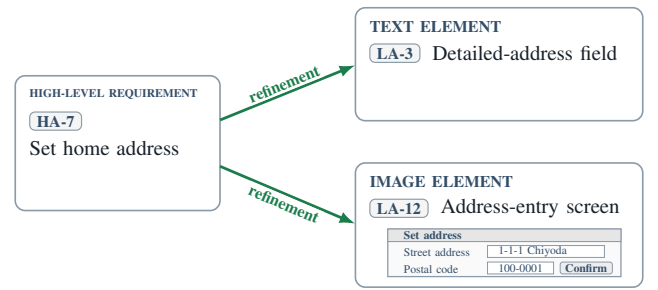

Figure~\ref{fig:tlr-example} illustrates this tracing task.
The high-level requirement states a user-facing capability, while the text and image elements in the low-level requirements refine parts of that capability.
TLR aims to recover both links from the possible HLR--LLR pairs.

The second challenge is adjusting the balance between precision and recall in industrial tracing tasks.
Requirements may share a screen or domain terms while specifying different controls or behavior.
Semantic similarity alone does not establish a valid trace link.
Based on our industrial experience, different tracing tasks within the same project may require different balances between precision and recall.
For a fixed classifier configuration, a yes-or-no output provides only one set of predicted links and no direct way to adjust this balance.

Our research goal is to improve automated link recovery for multimodal industrial requirements by reducing both missed links and incorrect predictions.
We also aim to help engineers balance missed links against incorrect links according to the needs of each tracing task.
We investigate how far a RAG-based inter-requirements TLR approach can support link recovery in multimodal industrial requirements.

\Ours extends the RAG-based design of Hey et al.~\cite{HeyEtAl2025RAGRequirementsTLR} in two ways.
Multimodal retrieval and a vision-language model (VLM) allow text and image elements to participate in candidate retrieval and relation classification.
The classifier also reports verbalized confidence alongside its binary decision.
Applying different confidence thresholds to the same responses can produce different subsets of the positive predictions.

The evaluation compares \ours with a text-only baseline that replicates the approach of Hey et al.~\cite{HeyEtAl2025RAGRequirementsTLR} in our execution environment.
We use non-public industrial datasets and the six public text-only datasets from their evaluation, which their replication package distributes as the \emph{req2req} collection~\cite{HeyEtAl2025Req2ReqArtifact}.

This paper makes the following contributions:

\begin{itemize}
  \item An empirical study of a replicated text-only RAG-based TLR approach on industrial test sets where most links involve images.
  \item \Ours, a multimodal extension of the same RAG-based design that retrieves candidates from both text and image elements and reports confidence for each predicted link.
  \item An evaluation of \ours on non-public industrial datasets and six public text-only datasets, covering confidence filtering and VLM sizes.
\end{itemize}

Sections~\ref{sec:related-work}, \ref{sec:methodology}, and~\ref{sec:empirical-study} present the background and related work, our approach, and the empirical study, respectively.
Sections~\ref{sec:discussion} and~\ref{sec:threats} discuss the findings and threats to validity, and Section~\ref{sec:conclusion} concludes the paper.

\section{Background and Related Work}\label{sec:related-work}

\subsection{Pre-LLM TLR Approaches}

Classical TLR treats each source artifact as a query and ranks target artifacts by textual similarity.
Hayes et al.~\cite{HayesEtAl2006CandidateLinks} evaluate vector-space and latent-semantic retrieval as support for human analysts.
Their study exemplifies a workflow in which automated retrieval generates candidates for analyst review.
Chen et al.~\cite{ChenEtAl2019S2Trace} combine ordered token patterns with document embeddings, while Hey et al.~\cite{HeyEtAl2021FTLR} map fine-grained requirement and code units before aggregating their relations.

\subsection{LiSSA and RAG-Based Inter-Requirements TLR}

Recent TLR studies~\cite{RodriguezEtAl2023PromptsMatter,BonnerEtAl2024LLMMBSE,HeyEtAl2025RAGRequirementsTLR,FuchssEtAl2025LiSSA} use LLMs to classify candidate trace links.
Rodriguez et al.~\cite{RodriguezEtAl2023PromptsMatter} show that prompt wording and task formulation can substantially change precision and recall in TLR\@.
In an industrial LLM-based TLR study, Bonner et al.~\cite{BonnerEtAl2024LLMMBSE} rank textual representations of requirements and model-based systems engineering artifacts for engineer review.
Fuch{\ss} et al.~\cite{FuchssEtAl2025LiSSA} present LiSSA, a general RAG-based framework for TLR\@.
LiSSA preprocesses artifacts into traceable elements, embeds the elements, retrieves target candidates for each source element, and uses an LLM to map candidate pairs before aggregating the mappings at the required granularity.
Its published evaluation covers requirements-to-code, documentation-to-code, and documentation-to-model trace tasks.
Hey et al.~\cite{HeyEtAl2025RAGRequirementsTLR} evaluate dense retrieval followed by zero-shot LLM classification on six benchmark datasets, including five datasets of HLRs and LLRs and one requirements-to-regulatory-codes dataset.
Their approach treats refinement and predecessor--successor relations as trace links.
Our method adopts LiSSA's artifact-and-element model and builds directly on the inter-requirements setting, link semantics, and public benchmarks evaluated by Hey et al.~\cite{HeyEtAl2025RAGRequirementsTLR}.

The simplest classifier prompts used by Hey et al.~\cite{HeyEtAl2025RAGRequirementsTLR} and Fuch{\ss} et al.~\cite{FuchssEtAl2025LiSSA} ask whether two artifacts are related.
The exact link semantics therefore depend on the artifact relation and oracle, rather than on semantic similarity alone.
Bonner et al.~\cite{BonnerEtAl2024LLMMBSE} describe linked artifacts in industrial TLR as items that refer to the same functionality or safety feature.
Preda et al.~\cite{PredaEtAl2024CoverageLLM} use LLMs to assess whether a given set of LLRs covers the aspects of an HLR, rather than to recover the pairwise links.

\subsection{Multimodal Requirements and Software Documents}

Vision-language models have been evaluated on adjacent software-document tasks.
Hassine~\cite{Hassine2025UMLVLM} evaluates diagram classification, element counting, boundary detection, and relationship extraction in UML use-case diagrams.
Fukuda and Tokumoto~\cite{FukudaTokumoto2026DiagramReview} combine rendered images with labels and connector data extracted from Office Open XML files, convert the recovered graph to PlantUML, and use it for design-document review.
These studies process visual software artifacts, but neither study recovers requirements trace links.

\subsection{Human Review and LLM Confidence}

Automated TLR has often been treated as support for a human analyst rather than as a replacement for expert judgment.
Hayes et al.~\cite{HayesEtAl2006CandidateLinks} distinguish link quality from the number of pairs that an analyst must inspect, and Cuddeback et al.~\cite{CuddebackEtAl2010HumanAnalysts} show that final results also depend on analyst verification.
In an interview study, Ruiz et al.~\cite{RuizEtAl2023WhyTrace} report accounts in which false positives made automated tracing unattractive and trace support needed to fit existing tools and processes.
These findings motivate reporting both recovery quality and the number of retained links.

Hey et al.~\cite{HeyEtAl2025RAGRequirementsTLR} and Fuch{\ss} et al.~\cite{FuchssEtAl2025LiSSA} classify each candidate pair using a binary yes-or-no output.
Verbalized confidence provides one possible signal for producing link sets with different balances between precision and recall.
Yang et al.~\cite{YangEtAl2024VerbalizedConfidence} define it as asking an LLM to express confidence as part of its generated response without requiring access to token logits.
They also show that the reliability of these scores varies across datasets and models and is strongly affected by prompt formulation.
A verbalized score should therefore not be treated as a calibrated probability without task-specific evidence.

\begin{table}[t]\centering
  \caption{Task Boundaries of Selected Related Studies}\label{tab:position}
  \setlength{\tabcolsep}{2.7pt}
  \begin{tabular}{@{}lccccc@{}} \toprule
    Study & TLR & \shortstack{Image\\Input} & \shortstack{Candidate\\Retrieval} & \shortstack{Industrial\\Data} & \shortstack{Confidence\\Filtering} \\ \midrule
    Hey et al. \cite{HeyEtAl2025RAGRequirementsTLR} & \cmark & & \cmark & & \\
    Fuch{\ss} et al. \cite{FuchssEtAl2025LiSSA} & \cmark & & \cmark & & \\
    Bonner et al. \cite{BonnerEtAl2024LLMMBSE} & \cmark & & \cmark & \cmark & \\
    Hassine \cite{Hassine2025UMLVLM} & & \cmark & & & \\
    Fukuda and Tokumoto \cite{FukudaTokumoto2026DiagramReview} & & \cmark & & \cmark & \\ \midrule
    This study & \cmark & \cmark & \cmark & \cmark & \cmark \\ \bottomrule
  \end{tabular}
\end{table}

Table~\ref{tab:position} positions the selected studies along five task dimensions.
The TLR studies~\cite{HeyEtAl2025RAGRequirementsTLR,FuchssEtAl2025LiSSA,BonnerEtAl2024LLMMBSE} operate on text, whereas the studies that process images directly~\cite{Hassine2025UMLVLM,FukudaTokumoto2026DiagramReview} do not recover trace links.
This study combines the two, and additionally filters the recovered links by confidence.

\section{Methodology}\label{sec:methodology}

We use a pretrained generative VLM to extend the prompt-based classification of Hey et al.~\cite{HeyEtAl2025RAGRequirementsTLR} to image inputs.
CLIP~\cite{RadfordEtAl2021CLIP} provides aligned image and text embeddings, but does not itself generate the trace decisions and verbalized confidence requested by our prompt.
Feeding its visual features into a text-only LLM requires an interface that aligns those features with the LLM's input representations.
For example, BLIP-2~\cite{LiEtAl2023BLIP2} trains a query transformer and a projection layer to connect a CLIP image encoder to a frozen LLM.
We use the visual interface already trained in the VLM, so images can be included in the classification prompt without developing and training this connection ourselves.

\subsection{Task and Representation of Requirement Elements}

Let \(\mathcal{H}=\{h_1,h_2,\ldots\}\) denote the HLR elements and \(\mathcal{L}=\{l_1,l_2,\ldots\}\) denote the LLR elements of one requirement set.
Each element contains either text or an image, and its modality is \(m(e)\in\{\mathrm{text},\mathrm{image}\}\).
Let \(Y\subseteq \mathcal{H}\times \mathcal{L}\) denote the target link set.
A pair in \(Y\) represents either a refinement relation or a predecessor--successor relation, following the inter-requirements task definition of Hey et al.~\cite{HeyEtAl2025RAGRequirementsTLR}.
The method produces a binary predicted link set \(\widehat{Y}\).
The output consists of element-level links from each HLR element to zero or more LLR elements.

\begin{figure*}[t]\centering
  \includegraphics[width=\textwidth]{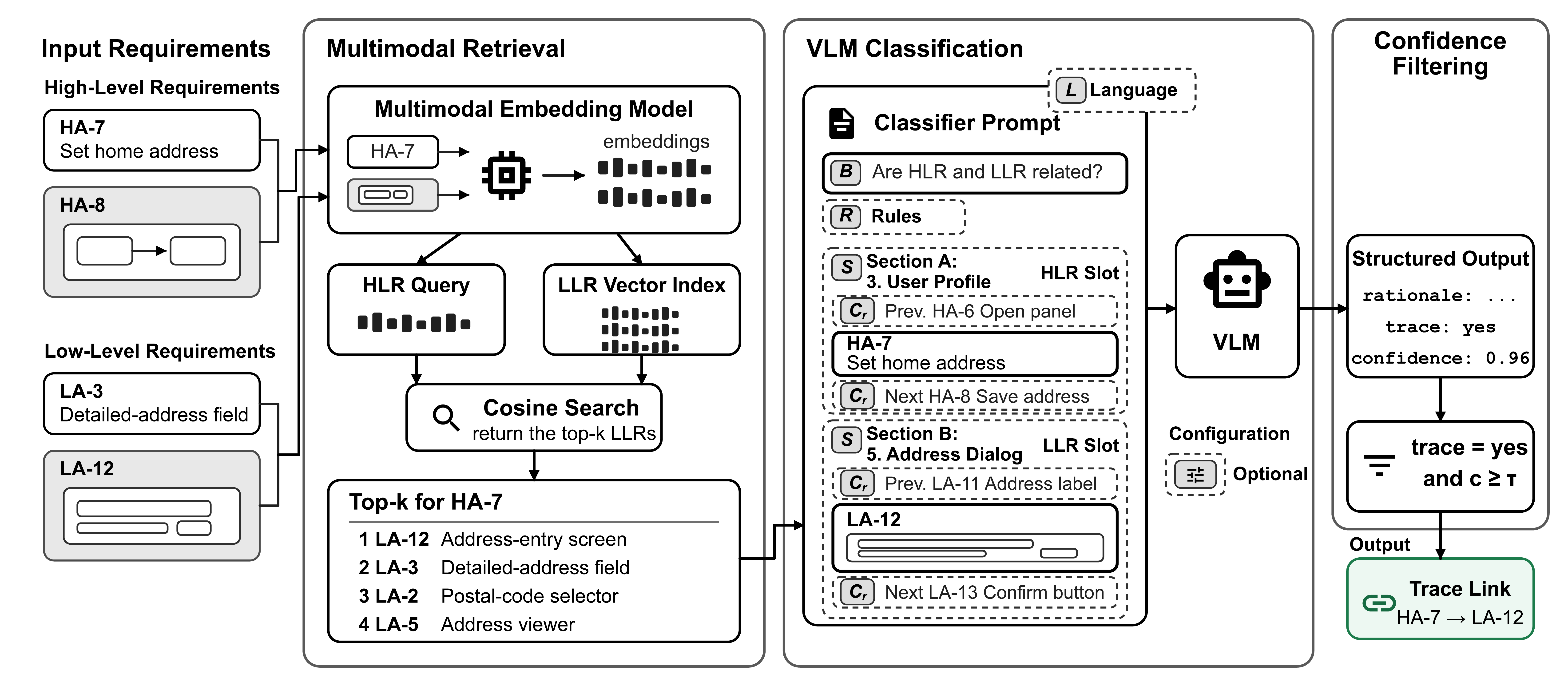}
  \caption{Illustrative multimodal TLR pipeline with the reference retrieval path and classifier input options.
    }\label{fig:method-pipeline}
\end{figure*}

Figure~\ref{fig:method-pipeline} presents our pipeline.
The text and image elements of an input requirement set are embedded with one multimodal model, and each HLR element retrieves its top-\(k\) LLR candidates (Section~\ref{sec:retrieval}).
A VLM classifies each HLR--candidate pair and returns a trace decision with a verbalized confidence (Section~\ref{sec:classification}).
Pairs with a positive decision whose confidence reaches the threshold become trace links.
The dashed boxes are optional variants of candidate allocation and classifier input (Section~\ref{sec:configurations}).

\subsection{Multimodal Candidate Retrieval}\label{sec:retrieval}

LiSSA defines top-\(k\) cosine retrieval from source elements to a vector store of target elements~\cite{FuchssEtAl2025LiSSA}.
Hey et al.\ use the same retrieval rule with whole requirements for inter-requirements TLR~\cite{HeyEtAl2025RAGRequirementsTLR}.
The \cfg{Base} setting (\(B\)) adapts this text-only rule by encoding the HLR and LLR elements with the same multimodal embedding model.
Let \(E(e)\) be the L2-normalized embedding of element \(e\).
For each requirement set, one index contains all LLR text and image embeddings, and each HLR element \(h\) queries that index, ranking the LLR elements by cosine similarity:
\begin{equation}
  s(h,l)=E(h)^{\mathsf{T}}E(l).
\end{equation}

Let \(\TOP_k(X;f)\) return the \(k\) elements of \(X\) with the largest values of \(f\).
The \cfg{Base} candidate set is
\begin{equation}
  \mathcal{C}_B(h)=\TOP_k(\mathcal{L};s(h,\cdot)).
\end{equation}

\cfg{Modality} (\(M\)) is our candidate-allocation extension and applies the same candidate count separately to each target modality.
Let \(\mathcal{L}_T\) and \(\mathcal{L}_I\) denote the text and image LLR elements.
Its candidate set is
\begin{equation}
  \mathcal{C}_M(h)=
    \TOP_k(\mathcal{L}_T;s(h,\cdot)) \cup
    \TOP_k(\mathcal{L}_I;s(h,\cdot)).
\end{equation}

\cfg{Base} (\(B\)) therefore passes \(k\) HLR--LLR element pairs per HLR element to the classifier, while \cfg{Modality} passes \(k\) text-target and \(k\) image-target pairs.
We fix \(k=4\) in the experiments, following Hey et al.~\cite{HeyEtAl2025RAGRequirementsTLR}, so \cfg{Base} returns four candidates per HLR element and \(B+M\) returns four text and four image candidates.
Section~\ref{sec:configuration-selection} compares \(M\) with the reference selection on the validation set.
The embeddings are used only to select candidates; the classifier receives the original text or image content.

\subsection{VLM Classification and Confidence Filtering}\label{sec:classification}

The classifier receives one retrieved HLR--LLR pair at a time.
A text element is inserted as a text message block, while an image element is provided as direct visual input.
The same input construction supports text--text, text--image, image--text, and image--image pairs.

The prompt asks the VLM to generate a rationale, the reasoning step of the prompt design of Hey et al.~\cite{HeyEtAl2025RAGRequirementsTLR}.
It then requests a \texttt{yes} or \texttt{no} decision in a \texttt{trace} tag and a score in a \texttt{confidence} tag.
This score is a verbalized confidence~\cite{YangEtAl2024VerbalizedConfidence}: it is stated by the model in its generated text rather than derived from its token probabilities.
The rationale is retained in the raw response but is not used to construct the final links.
The trace parser accepts a case-insensitive \texttt{yes} or \texttt{no} value inside the trace tag.
A missing or unrecognized trace value is treated as unrelated, following the positive-link construction rule used by Hey et al.~\cite{HeyEtAl2025RAGRequirementsTLR}.
The requested confidence range is \([0,1]\).
A missing confidence tag is assigned \(1.0\) so that the parsed trace decision is retained at every threshold, while a nonnumeric, non-finite, or out-of-range value indicates a malformed response and is assigned \(0.0\).

Let \(\mathcal{C}(h)\) denote the candidate set produced by the selected configuration.
Let \(d(h,l)\in\{0,1\}\) be the parsed trace decision and \(c(h,l)\) be its confidence.
Without confidence filtering, the predicted links are
\begin{equation}
  \widehat{Y}_{\mathrm{raw}} =
    \{(h,l)\mid l\in \mathcal{C}(h)\land d(h,l)=1\}.
\end{equation}

We refer to \(\widehat{Y}_{\mathrm{raw}}\) as the unfiltered link set, labeled \emph{No Filter} in tables and figures.
It is constructed from the parsed trace decision alone.

At threshold \(\tau\), the accepted links are
\begin{equation}
  \widehat{Y}_\tau =
    \{(h,l)\mid l\in \mathcal{C}(h)\land d(h,l)=1\land c(h,l)\geq\tau\}.
\end{equation}

Each filtering condition produces a binary set of links.
The method definition leaves \(\tau\) as a parameter, and Section~\ref{sec:empirical-study} describes how it is selected before test-set evaluation.

\subsection{Configuration Options}\label{sec:configurations}

The method supports several configurations on the candidate retrieval and the reference classifier input.
Table~\ref{tab:method-configurations} defines the reference configuration \(B\) (\cfg{Base}) and five single-factor variants.
Each variant changes one part of \(B\), while the model, decoding settings, output format, and parsing rules remain fixed.

\begin{table}[t]\centering
  \caption{Configuration Options}\label{tab:method-configurations}
  \begin{tabular}{@{}lll@{}}\toprule
    ID & Changed part & Setting \\ \midrule
    \(B\)       & Reference & Shared top-\(k\) and generic English prompt \\
    \(B+R\)     & Prompt instruction & Trace-link and decision rules \\
    \(B+L\)     & Prompt instruction & Instruction in artifact language \\
    \(B+C_r\)   & Prompt content & \(r\) preceding and \(r\) following elements \\
    \(B+S\)     & Prompt content & Section headings \\
    \(B+M\)     & Candidate retrieval & \(k\) text and \(k\) image candidates \\ \bottomrule
  \end{tabular}%
\end{table}

Configuration \(B\) (\cfg{Base}) uses the reference retrieval rule \(\mathcal{C}_B\) and a generic English instruction that provides only the two focal elements.
It follows the generic trace question and rationale structure used by Hey et al.~\cite{HeyEtAl2025RAGRequirementsTLR}, extended with the confidence field required by \ours.
Configuration \(B+R\) (\cfg{Rules}) adds rules that define refinement and predecessor--successor as the two target relation types and state that shared context alone does not establish a link.
Configuration \(B+L\) (\cfg{Language}) uses instructions in the language of the requirement artifacts while retaining the same task and output tags.
Configuration \(B+C_r\) (\cfg{Context}) adds the available \(r\) preceding and \(r\) following elements from the same source document for each focal element.
The neighboring elements retain their original order and text or image modalities, and the context is supplied only to the classifier.
We evaluate this configuration at the fixed radius \(r=4\), adding up to four preceding and four following elements for each focal element while leaving the retrieval candidate count unchanged.
Configuration \(B+S\) (\cfg{Section}) adds the stored section headings of the two focal elements to the classifier prompt.
Configuration \(B+M\) (\cfg{Modality}) uses \(\mathcal{C}_M(h)\) while retaining the reference classifier input.

\section{Empirical Study}\label{sec:empirical-study}

\begin{table*}[t]\centering
  \caption{Dataset Characteristics}\label{tab:datasets}
  \begin{tabular*}{\textwidth}{@{\extracolsep{\fill}}ll*{10}{r}@{}} \toprule
    Group & Dataset & \multicolumn{4}{c}{Requirement Elements} &
    \multicolumn{5}{c}{Gold Links} &
    \multicolumn{1}{c}{\# of Gold Links} \\
    \cmidrule(lr){3-6}\cmidrule(lr){7-11}\cmidrule(l){12-12}
    & & \multicolumn{2}{c}{HLR} & \multicolumn{2}{c}{LLR} &
    \multicolumn{1}{c}{Text--Text} & \multicolumn{4}{c}{Image-Involving} & \\
    \cmidrule(lr){3-4}\cmidrule(lr){5-6}\cmidrule(lr){7-7}
    \cmidrule(lr){8-11}
    & & Text & Image & Text & Image & TT & TI & IT & II & Subtotal & \\ \midrule
    \multirow{4}{*}{Industrial}
      & \SetA & 57 & 4 & 267 & 21 & 145 & 8 & 30 & 2 & 40 (21.6\%) & 185 \\
      & \SetB & 39 & 5 & 180 & 11 & 61 & 14 & 10 & 8 & 32 (34.4\%) & 93 \\
      & \SetC & 7 & 12 & 227 & 20 & 0 & 0 & 50 & 28 & 78 (100.0\%) & 78 \\ \cmidrule(l){2-12}
      & Total              & 103 & 21 & 674 & 52 & 206 & 22 & 90 & 38 & 150 (42.1\%) & 356 \\ \midrule
    \multirow{7}{*}{Public}
      & CCHIT      & 116 & 0 & 1,064 & 0 & 587 & 0 & 0 & 0 & 0 (0.0\%) & 587 \\
      & CM1-NASA   & 22  & 0 & 53    & 0 & 45  & 0 & 0 & 0 & 0 (0.0\%) & 45 \\
      & Dronology  & 99  & 0 & 211   & 0 & 220 & 0 & 0 & 0 & 0 (0.0\%) & 220 \\
      & GANNT      & 17  & 0 & 69    & 0 & 68  & 0 & 0 & 0 & 0 (0.0\%) & 68 \\
      & MODIS      & 19  & 0 & 49    & 0 & 41  & 0 & 0 & 0 & 0 (0.0\%) & 41 \\
      & WARC       & 63  & 0 & 89    & 0 & 136 & 0 & 0 & 0 & 0 (0.0\%) & 136 \\ \cmidrule(l){2-12}
      & Total      & 336 & 0 & 1,535 & 0 & 1,097 & 0 & 0 & 0 & 0 (0.0\%) & 1,097 \\ \bottomrule
  \end{tabular*}%
\end{table*}

Industrial requirements commonly combine text with graphical elements~\cite{FranchEtAl2023RequirementsPractice}, but recent inter-requirements TLR evaluation uses text artifacts~\cite{HeyEtAl2025RAGRequirementsTLR}.
Existing results do not show how much of an industrial task falls outside text-only input or whether direct image processing can recover these links while retaining text-only performance.
Different tracing activities can prioritize precision and recall differently, so one fixed link set may not fit each activity.
Because the feasible VLM size is bounded by available hardware, we examine how much link-recovery quality smaller checkpoints retain.
We study these gaps through four research questions (RQs):
\begin{itemize}
\item \textbf{\RQ{1}: }
How much of the industrial test oracle is accessible to a text-only approach, and how does the Hey et al.\ baseline perform on this scope and on all link types?
\item \textbf{\RQ{2}: }
Does \ours improve recovery over all industrial link types while retaining text-only recovery?
\item \textbf{\RQ{3}: }
Can confidence filtering provide useful link sets for tracing tasks with different precision and recall priorities?
\item \textbf{\RQ{4}: }
How much link-recovery quality is lost when a smaller VLM is used under tighter deployment constraints?
\end{itemize}

For each RQ, the following subsections specify the input data, the executed approach, the oracle scope, and how the results are combined.

\subsection{Experimental Setup}

\subsubsection{Datasets and Oracle Scopes}\label{sec:datasets}

The public data comprise six English text-to-text benchmark datasets evaluated by Hey et al.~\cite{HeyEtAl2025RAGRequirementsTLR} and distributed in their replication package~\cite{HeyEtAl2025Req2ReqArtifact}.
We use the gold-link oracle distributed with each public dataset.
These datasets serve to evaluate text-only compatibility (\RQ{2}).

The industrial data comprise three non-overlapping Japanese requirement sets from different parts of one system.
We denote the three industrial sets as \SetA, \SetB, and \SetC.
\SetA serves as the validation set, denoted \SetValid, and the union of \SetB and \SetC serves as the test set, denoted \SetTest.
We compare configurations and select confidence thresholds on \SetValid, then freeze these choices before evaluating \SetTest.

Each HLR or LLR document is represented as an ordered sequence of paragraph and image elements.
An image element can be a figure, an image-based table, or a consolidated workflow.
We semi-manually convert the industrial documents to the HLR, LLR, and oracle organization of req2req, extended with image elements.
Paragraphs and visual content become text and direct image elements in their original document order.
We manually inspect the converted inputs and retain section headings as metadata.

The gold-link oracles for the industrial sets were annotated for this study by two practitioners with 10 and 13 years of software-development experience, respectively.
Each requirement set was labeled by one practitioner.
Each industrial oracle assigns a binary decision to every HLR--LLR element pair.
A positive pair is a gold link representing a refinement or predecessor--successor relation, following the relation family used by Hey et al.~\cite{HeyEtAl2025RAGRequirementsTLR}.

In Table~\ref{tab:datasets}, TT (Text-to-Text pair), TI (Text-to-Image pair), IT (Image-to-Text pair), and II (Image-to-Image pair) give the source and target modalities in that order.
The Image-Involving subtotal sums TI, IT, and II, with its share of all gold links shown in parentheses.
The rightmost \# of Gold Links column includes all these four types.
It shows that 150 of the 356 industrial links (42.1\%) involve at least one image.
Note that every gold link involves at least one image and no text--text gold link exists in \SetC, which illustrates why TLR must handle image elements.

For each industrial set, the full input contains every text and image element.
For each test subset, the derived text-only input is obtained by removing image elements.
A superscript \(\text{text-only}\) marks this operation, as in \SetBTextOnly and \SetCTextOnly.
We write \SetTestTextOnly for their combined results.
We evaluate text-only predictions against the TT oracle scope and evaluate task-wide coverage against the full oracle.
The image-involving oracle scope is the disjoint union of TI, IT, and II.

\subsubsection{Metrics}

Let TP, FP, and FN denote true-positive, false-positive, and false-negative predicted links.
We compute precision, recall, and \(F_\beta\) as follows:
\begin{equation}
  P\!=\!\frac{\mathrm{TP}}{\mathrm{TP}+\mathrm{FP}},~
  R\!=\!\frac{\mathrm{TP}}{\mathrm{TP}+\mathrm{FN}},~
  F_\beta\!=\!(1+\beta^2)\frac{PR}{\beta^2P+R}.
\end{equation}

When no positive link is predicted, the precision denominator in Eq.~(6) is zero, so precision is reported as undefined~\mbox{(--)}.
When gold links exist but none is recovered, the recall numerator in Eq.~(6) is zero and its denominator is positive, so recall is zero.
We also report \(F_\beta\) as zero in this case.
For a separately reported modality subset with no gold link, the recall denominator in Eq.~(6) is zero, so recall is reported as undefined~\mbox{(--)}.
We report \(F_\beta\) as not applicable~\mbox{(--)} for these subsets.
Following binary-classifier evaluation~\cite{SaitoRehmsmeier2015PRPlot} and industrial TLR evaluation~\cite{BonnerEtAl2024LLMMBSE}, we use \(\beta\in\{0.5,1,2\}\).
Values below one give more weight to precision, while values above one give more weight to recall~\cite{ManningEtAl2008IR}.
Metrics for a combination of sets or oracle scopes instead sum their TP, FP, and FN counts before calculation.
We use precision, recall, and \(F_\beta\) because negative pairs dominate the Cartesian product.

Text-only coverage is the proportion of gold links in the full oracle whose two endpoints are text elements.
The prediction set of the text-only baseline (Section~\ref{sec:approaches}) is evaluated against the text--text oracle subset and against the full industrial oracle, and the separate text-only run of \ours is evaluated against the same text--text subset.

\subsubsection{Compared Approaches}\label{sec:approaches}

\HeyRep is our executable text-only replication of the inter-requirements approach published by Hey et al.~\cite{HeyEtAl2025RAGRequirementsTLR}.
It retains the published pipeline but uses locally hosted open-weight models, Qwen3-Embedding-8B and Mistral Small 3.2 24B, instead of the proprietary models because the industrial requirements could not be sent to external services.
The replication uses the latest repository revisions available when the experiments began.%
\footnote{Revisions: Qwen3-Embedding-8B \revision{1d8ad4ca9b3d} and Mistral Small 3.2 24B \revision{95a6d26c4bfb}.}
Qwen3-Embedding-8B provides multilingual retrieval~\cite{ZhangEtAl2025Qwen3Embedding}, and Mistral Small 3.2 24B fits one 80~GB H100 in bfloat16~\cite{MistralAI2025Small32,NVIDIA2025MistralSmall32NIM}.
The notation \(\cdot_{\mathrm{rep}}\) marks results produced by this executable replication rather than values copied from the published paper.

\Ours uses a multimodal embedding model to retrieve text and image LLR elements for each HLR element.
A Gemma 4 VLM classifies each retrieved pair and generates a binary decision and confidence score.
For the industrial text-only comparison, \ours uses the reference classifier input and retrieves the four highest-similarity text candidates because configuration \(M\) requires both target modalities.
The MMEB-V2 benchmark ranks Qwen3-VL-Embedding-8B first among the evaluated multimodal embedding models~\cite{TIGERLab2026MMEBLeaderboard,LiEtAl2026Qwen3VLEmbedding}.
Under the one-H100 hardware constraint, the Vision Arena benchmark identifies Gemma 4 31B as the highest-ranked model that fits the available hardware~\cite{ArenaAI2026VisionLeaderboard,GemmaTeam2026Gemma4}.
The multimodal runs also use the latest repository revisions available when the experiments began.%
\footnote{Revisions: Qwen3-VL-Embedding-8B \revision{2c4565515e0f}; Gemma 4 E2B \revision{3e22461f65e8}, 12B \revision{707f0a3b8a3c}, and 31B \revision{842da3794eaa}.}
The primary VLM and all \RQ{4} model variants are fixed before \SetTest is evaluated.

Table~\ref{tab:configurations} records the published configuration of Hey et al.~\cite{HeyEtAl2025RAGRequirementsTLR} and the primary settings executed in this study.
The Prompt row lists the fields requested from the classifier: the reasoning step, the trace decision, and, for \ours, the confidence score (Conf.).
The Parser row lists the fields used to construct links.
Before using the replication as the industrial baseline, we compared it with the public results reported by Hey et al.~\cite{HeyEtAl2025RAGRequirementsTLR} in Table~\ref{tab:replication-check}.
The arithmetic mean \(F_1\) values across the six datasets are .451 and .449 for the reported and replicated results.
We therefore use the executed replication as the proxy for Hey et al.\ in the remaining comparisons, label it \emph{Hey et al.} in the tables, and refer to it as the baseline in the text.

\begin{table}[t]\centering
  \caption{Published and Executed Configurations}\label{tab:configurations}
  \setlength{\tabcolsep}{1.5pt}
  \begin{tabular}{@{}llll@{}} \toprule
    Setting & Hey et al.~\cite{HeyEtAl2025RAGRequirementsTLR} & \HeyRep & \OurS \\ \midrule
    Embedding & text-emb.-3-large & Qwen3-Emb.-8B & Qwen3-VL-Emb.-8B \\
    Classifier & GPT-4o & Mistral 24B & Gemma 4 31B \\
    Prompt & Reasoning \!+\! Trace & Reasoning \!+\! Trace & Reasoning \!+\! Trace \!+\! Conf. \\
    Parser & Trace & Trace & Trace \!+\! Conf. \\
    Candidates & Top \(k\) & Top \(k\) & Top \(k\) per Modality \\
    Temperature & 0 & 0 & 0 \\ \bottomrule
  \end{tabular}%
\end{table}

\begin{table}[t]\centering
  \caption{Published and Replicated Public Results}\label{tab:replication-check}
  \begin{tabular}{@{}llrrr@{}}\toprule
    Dataset & Result & $P$ & $R$ & $F_1$ \\ \midrule
    \multirow{2}{*}{CCHIT}
      & Hey et al. & .367 & .138 & .200 \\
      & \HeyRep & .319 & .141 & .196 \\ \addlinespace[2pt]
    \multirow{2}{*}{CM1-NASA}
      & Hey et al. & .458 & .600 & .519 \\
      & \HeyRep & .403 & .600 & .482 \\ \addlinespace[2pt]
    \multirow{2}{*}{Dronology}
      & Hey et al. & .512 & .655 & .575 \\
      & \HeyRep & .466 & .645 & .541 \\ \addlinespace[2pt]
    \multirow{2}{*}{GANNT}
      & Hey et al. & .607 & .544 & .574 \\
      & \HeyRep & .542 & .471 & .504 \\ \addlinespace[2pt]
    \multirow{2}{*}{MODIS}
      & Hey et al. & .500 & .171 & .255 \\
      & \HeyRep & .500 & .341 & .406 \\ \addlinespace[2pt]
    \multirow{2}{*}{WARC}
      & Hey et al. & .537 & .640 & .584 \\
      & \HeyRep & .475 & .691 & .563 \\ \midrule
    \multirow{2}{*}{Mean}
      & Hey et al. & .497 & .458 & .451 \\
      & \HeyRep & .451 & .482 & .449 \\ \bottomrule
  \end{tabular}
\end{table}

\subsubsection{Evaluation Procedure}

Each configuration uses temperature zero and a maximum output length of 1,024.
The unquantized bfloat16 runs use one 80~GB NVIDIA H100 with PyTorch 2.8, CUDA 12.9, and Transformers 5.14.

\subsection{Configuration Selection}\label{sec:configuration-selection}

We compare \(B\) with the single-factor configurations \(B+R\), \(B+L\), \(B+C_4\), \(B+S\), and \(B+M\) of Table~\ref{tab:method-configurations} on \SetValid.
Configuration \(B+M\) selects four text and four image candidates.
The other configurations retain the shared top-4 candidates and change one classifier input factor.
For each configuration, we apply confidence thresholds from zero to one in increments of 0.05 to its stored classifier responses and record its maximum overall \(F_1\).
If several thresholds give the same maximum, we use the higher threshold.
Because \(F_1\) weights precision and recall equally, we select the configuration with the highest maximum \(F_1\).
Using its responses, we select the thresholds that maximize \(F_{0.5}\), \(F_1\), and \(F_2\), again using the higher threshold for tied maxima.

\begin{table}[t]\centering
  \caption{Configuration Comparison for \OurS on \SetValid\\at Each Row's Best \(F_1\) Threshold}\label{tab:validation-selection}
  \begin{tabular}{@{}lrrrrr@{}} \toprule
    Configuration & $\tau$ at max. $F_1$ & $P$ & $R$ & $F_1$ & Candidate recall \\ \midrule
    \(B\)      & 1.00 & .694 & .368 & .481 & .438 \\
    \(B+R\)    & 1.00 & .817 & .314 & .453 & .438 \\
    \(B+L\)    & 1.00 & .661 & .389 & .490 & .438 \\
    \(B+C_4\)  & .95 & .411 & .422 & .416 & .438 \\
    \(B+S\)    & 1.00 & .685 & .341 & .455 & .438 \\
    \(B+M\)    & 1.00 & .667 & .422 & .517 & .492 \\ \bottomrule
  \end{tabular}
\end{table}

Table~\ref{tab:validation-selection} shows, for each configuration, its best point on the threshold grid and its candidate recall.
The \(B+M\) configuration obtains the highest maximum overall \(F_1\), .517 at \(\tau=1.00\), compared with .481 for \(B\).
The classifier input variants obtain at most .490.
Candidate recall, the proportion of gold links present in the retrieved set, increases from .438 to .492.
For text--image and image--text links, it increases from zero to .500 and .200, respectively.
We therefore fix the configuration to \(B+M\) before evaluating \SetTest.

\begin{table*}[t]\centering
  \caption{Detailed Results of \OurS on Industrial and Public Datasets}\label{tab:ours-by-set}
  \setlength{\tabcolsep}{3.5pt}
  \renewcommand{\arraystretch}{1.00}
  \begin{tabular*}{\textwidth}{@{\extracolsep{\fill}}lll*{15}{r}@{}}\toprule
    & & & \multicolumn{3}{c}{Overall} & \multicolumn{3}{c}{TT} &
    \multicolumn{3}{c}{TI} & \multicolumn{3}{c}{IT} & \multicolumn{3}{c}{II} \\
    \cmidrule(lr){4-6}\cmidrule(lr){7-9}\cmidrule(lr){10-12}
    \cmidrule(lr){13-15}\cmidrule(l){16-18}
    Group & Dataset & Output &
    $P$ & $R$ & $F_1$ & $P$ & $R$ & $F_1$ & $P$ & $R$ & $F_1$ &
    $P$ & $R$ & $F_1$ & $P$ & $R$ & $F_1$ \\ \midrule
    \multirow{14}{*}{Industrial}
      & \multirow{4}{*}{\SetA}
      & No Filter & .288 & .486 & .361 & .398 & .538 & .457 &
        .043 & .500 & .079 & .429 & .200 & .273 & .200 & 1.00 & .333 \\
      & & Max. $F_{0.5}$ & .667 & .422 & .517 & .702 & .455 & .552 &
        .364 & .500 & .421 & .750 & .200 & .316 & .500 & 1.00 & .667 \\
      & & Max. $F_1$ & .667 & .422 & .517 & .702 & .455 & .552 &
        .364 & .500 & .421 & .750 & .200 & .316 & .500 & 1.00 & .667 \\
      & & Max. $F_2$ & .419 & .476 & .446 & .507 & .524 & .515 &
        .103 & .500 & .170 & .462 & .200 & .279 & .250 & 1.00 & .400 \\ \addlinespace
      & \multirow{4}{*}{\SetB}
      & No Filter & .164 & .430 & .237 & .177 & .361 & .238 &
        .123 & .714 & .211 & .000 & .000 & .000 & .400 & 1.00 & .571 \\
      & & Max. $F_{0.5}$ & .296 & .312 & .304 & .500 & .197 & .282 &
        .209 & .643 & .316 & .000 & .000 & .000 & .471 & 1.00 & .640 \\
      & & Max. $F_1$ & .296 & .312 & .304 & .500 & .197 & .282 &
        .209 & .643 & .316 & .000 & .000 & .000 & .471 & 1.00 & .640 \\
      & & Max. $F_2$ & .200 & .409 & .269 & .238 & .328 & .276 &
        .147 & .714 & .244 & .000 & .000 & .000 & .400 & 1.00 & .571 \\ \addlinespace
      & \SetBTextOnly & No Filter & .176 & .237 & .202 & .176 & .361 & .237 &
        -- & .000 & .000 & -- & .000 & .000 & -- & .000 & .000 \\ \addlinespace
      & \multirow{4}{*}{\SetC}
      & No Filter & .235 & .346 & .280 & .000 & -- & -- &
        .000 & -- & -- & .190 & .160 & .174 & .514 & .679 & .585 \\
      & & Max. $F_{0.5}$ & .426 & .256 & .320 & .000 & -- & -- &
        .000 & -- & -- & .350 & .140 & .200 & .722 & .464 & .565 \\
      & & Max. $F_1$ & .426 & .256 & .320 & .000 & -- & -- &
        .000 & -- & -- & .350 & .140 & .200 & .722 & .464 & .565 \\
      & & Max. $F_2$ & .321 & .346 & .333 & .000 & -- & -- &
        .000 & -- & -- & .250 & .160 & .195 & .594 & .679 & .633 \\ \addlinespace
      & \SetCTextOnly & No Filter & .000 & .000 & .000 & .000 & -- & -- &
        -- & -- & -- & -- & .000 & .000 & -- & .000 & .000 \\ \midrule
    \multirow{6}{*}{Public}
      & CCHIT & No Filter & .333 & .152 & .208 & .333 & .152 & .208 &
        -- & -- & -- & -- & -- & -- & -- & -- & -- \\
      & CM1-NASA & No Filter & .464 & .711 & .561 & .464 & .711 & .561 &
        -- & -- & -- & -- & -- & -- & -- & -- & -- \\
      & Dronology & No Filter & .412 & .636 & .500 & .412 & .636 & .500 &
        -- & -- & -- & -- & -- & -- & -- & -- & -- \\
      & GANNT & No Filter & .492 & .426 & .457 & .492 & .426 & .457 &
        -- & -- & -- & -- & -- & -- & -- & -- & -- \\
      & MODIS & No Filter & .500 & .268 & .349 & .500 & .268 & .349 &
        -- & -- & -- & -- & -- & -- & -- & -- & -- \\
      & WARC & No Filter & .454 & .684 & .545 & .454 & .684 & .545 &
        -- & -- & -- & -- & -- & -- & -- & -- & -- \\ \bottomrule
  \end{tabular*}
\end{table*}

\subsection{\RQ{1}: Text-Only Coverage}\label{sec:rq1}

\subsubsection{Motivation}
The first challenge in Section~\ref{sec:introduction} concerns image elements that text-only input cannot represent.
Hey et al.'s public benchmark results leave open how much of the industrial task their approach can support.
Assessing this gap requires distinguishing links involving excluded image elements from recovery errors on the available text elements.

\subsubsection{Design}
We derive \SetBTextOnly and \SetCTextOnly by removing image elements from both the HLR and LLR documents in \SetB and \SetC while retaining the identifiers and order of the text elements.
We run the baseline separately on the two derived inputs.
For each HLR text element, the approach retrieves the four highest-similarity LLR text candidates and classifies the resulting pairs.
We retain each prediction set and evaluate it under two oracle scopes.
The TT scope contains only gold links whose source and target are both text elements and therefore measures the part of the task available to the text-only input.
The full scope contains TT, TI, IT, and II gold links.
The predictions remain unchanged in this second evaluation, so each unrecovered image-involving gold link contributes a false negative.
We also count the TI, IT, and II gold links in the full oracle to measure how much of the task lies outside the text-only input.

\subsubsection{Results and Discussion}
\SetTest contains 171 gold links, of which 110 (64.3\%) involve at least one image element.
The baseline recovers 17 of the 61 TT links, giving \(P=.135\), \(R=.279\), and \(F_1=.182\).
Against all 171 links, the same 17 correct predictions give \(R=.099\) and \(F_1=.114\).
\SetC contains no text--text gold link, so its task over all link types lies entirely outside the baseline input.

\RQAnswer{1}{The text-only baseline recovers 27.9\% of the links accessible to its text-only input, but only 9.9\% of the full industrial oracle. A text-only approach can access only 35.7\% of the task because the remaining \emph{64.3\% lies outside its text-only input}. Most industrial links therefore require image-aware input.}

\subsection{\RQ{2}: Multimodal Recovery and Text-Only Compatibility}\label{sec:rq2}

\subsubsection{Motivation}
To address the coverage gap examined in \RQ{1}, the extension needs to recover correct links involving images.
Changes to retrieval and classification may also affect its performance on textual requirements.
We therefore assess whether the extended pipeline improves industrial link recovery while preserving the text-only capability of the original approach.

\subsubsection{Design}
We conduct three comparisons.
First, we run \ours with the fixed \(B+M\) configuration and the primary 31B VLM on the full text and image inputs of \SetB and \SetC.
Besides the unfiltered link set, we form filtered link sets by applying the thresholds selected on \SetValid for \(F_{0.5}\), \(F_1\), and \(F_2\) to the same responses.
We evaluate each link set against the full oracle, separate the results into TT, TI, IT, and II links, and compare them with the baseline predictions from \RQ{1} evaluated against the same full oracle.

Second, we evaluate text-only compatibility on the industrial data.
We run \ours separately on \SetBTextOnly and \SetCTextOnly and compare its unfiltered predictions with the baseline predictions from \RQ{1}.
Because \cfg{Modality} (\(M\)) requires both target modalities, this run of \ours uses \cfg{Base} (\(B\)) and retrieves the four highest-similarity text candidates.
We evaluate the prediction sets against the TT oracle to compare text-only recovery and against the full oracle to retain the task-coverage view used in \RQ{1}.
We also extract the TT results from the unfiltered full-input runs of \ours to examine whether the presence of image elements changes text--text recovery.

Third, we run both approaches without filtering on each of the six public text-to-text datasets, again using \cfg{Base} with shared top-4 retrieval for \ours.
For the combined public result, we sum TP, FP, and FN across the six datasets before calculating the measures.
This count-based result is separate from the arithmetic mean in Table~\ref{tab:replication-check}, which follows the aggregation reported by Hey et al.

\subsubsection{Results and Discussion}
Table~\ref{tab:ours-by-set} reports the results.
For the full industrial inputs, Max. \(F_\beta\) denotes the link set at the threshold selected by that measure on \SetValid.
For the derived text-only rows, Overall uses the full oracle, while TT reports the input-accessible oracle scope.
Because the thresholds were selected on \SetValid, the Max. \(F_\beta\) rows of \SetA are in-sample values and are not comparable to the corresponding rows of \SetB and \SetC.

Table~\ref{tab:scope-comparison} organizes the comparison by its four oracle scopes.
The Industrial All and Image-Involving rows use the link set selected by both \(F_{0.5}\) and \(F_1\), while the two text-only rows use the unfiltered link sets.

\begin{table}[t]\centering
  \caption{Hey et al. and \OurS by Oracle Scope}\label{tab:scope-comparison}
  \begin{tabular}{@{}lrrrrrr@{}}\toprule
    & \multicolumn{3}{c}{Hey et al.} & \multicolumn{3}{c}{\OurS} \\
    \cmidrule(lr){2-4}\cmidrule(l){5-7}
    Scope & $P$ & $R$ & $F_1$ & $P$ & $R$ & $F_1$ \\ \midrule
    Industrial All & .135 & .099 & .114 & .338 & .287 & .310 \\
    Industrial \SetTestTextOnly & .135 & .279 & .182 & .152 & .361 & .214 \\
    Industrial Image-Involving & -- & .000 & .000 & .322 & .336 & .329 \\
    Public Combined & .427 & .357 & .389 & .410 & .359 & .383 \\ \bottomrule
  \end{tabular}
\end{table}

Across \SetTest, \ours recovers 49 of 171 gold links with \(P=.338\), \(R=.287\), and \(F_1=.310\).
Compared with 17 correctly recovered links and \(F_1=.114\) for the baseline, this is an increase of 32 correctly recovered links and \(2.71\times\) the \(F_1\).
These results compare the complete executed systems, including changes to input modalities, embedding model, classifier, prompt, parser, and confidence filtering.
The candidate counts also differ: the baseline retrieves four text candidates per text HLR, while \ours retrieves four text and four image candidates per HLR.
The \(\tau=1.00\) link set recovers 37 of the 110 image-involving links: 9 of 14 text--image, 7 of 60 image--text, and 21 of 36 image--image links.
The 16 links in the two cross-modal directions were outside the text-only input.
On \SetTestTextOnly, \ours raises industrial \(F_1\) from .182 to .214 and obtains .383 on the public data compared with .389 for the baseline.
On \SetTest, the unfiltered full-input runs achieve TT \(F_1\) of .215 versus .214 for the text-only runs, suggesting that the pipeline can extend recovery to image-involving links while retaining its text--text capability.

\RQAnswer{2}{\textbf{(1)} \Ours recovers \emph{49 of 171 industrial gold links}, with precision .338 and recall .287.
This is 188\% more correctly recovered links than the text-only baseline, and it includes 33.6\% of the image-involving gold links.
\textbf{(2)} Its text-only capability is retained.
Industrial \(F_1\) improves, and public \(F_1\) changes by less than .01.}

\subsection{\RQ{3}: Confidence Trade-Offs}\label{sec:rq3}

\subsubsection{Motivation}
The second challenge in Section~\ref{sec:introduction} concerns tracing tasks with different precision and recall priorities.
These priorities require a way to adjust the classifier's predicted link set, but verbalized confidence varies in reliability across tasks.
We therefore examine whether thresholds selected on \SetValid provide useful precision--recall balances on \SetTest.

\subsubsection{Design}
We classify each retrieved pair once with the \(B+M\) configuration selected on \SetValid and reuse the stored responses throughout this experiment.
For each threshold from zero to one in increments of 0.05, we retain a positive decision only when its confidence is at least the threshold.
We compute precision, recall, \(F_{0.5}\), \(F_1\), and \(F_2\) for every threshold on \SetValid.
For each \(F_\beta\), we select the threshold that maximizes that measure, choosing the higher threshold when several values tie, and apply it without adjustment to the stored responses from \SetB and \SetC.
We also calculate the complete test-set threshold curves to describe how precision, recall, and \(F_1\) change, but these curves do not alter the thresholds selected on \SetValid.

\subsubsection{Results and Discussion}
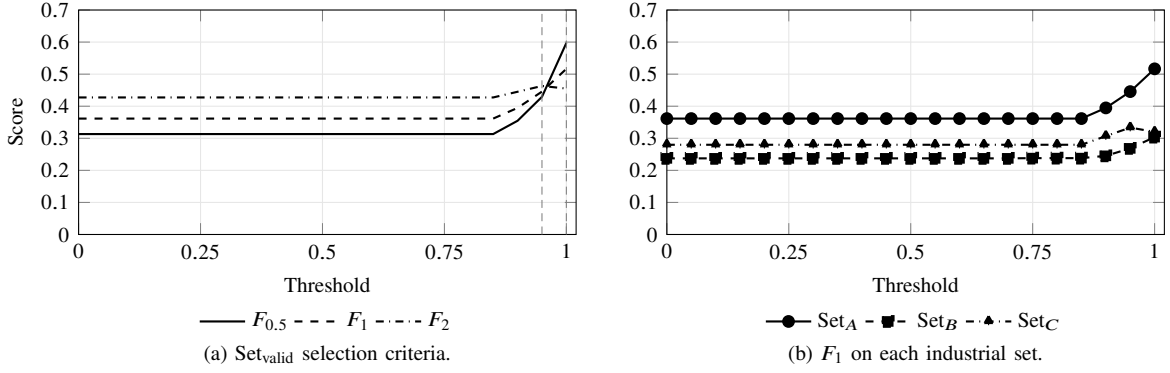
\begin{figure*}[t]\centering
\begin{tikzpicture}
\begin{groupplot}[
  group style={group size=2 by 1,horizontal sep=1.2cm},
  width=0.45\textwidth,
  height=4.55cm,
  xmin=0,
  xmax=1.02,
  ymin=0,
  ymax=0.7,
  xlabel={Threshold},
  ylabel={Score},
  xtick={0,.25,.5,.75,1},
  ytick={0,.1,.2,.3,.4,.5,.6,.7},
  grid=major,
  grid style={gray!20},
  tick label style={font=\footnotesize},
  label style={font=\footnotesize},
  title style={font=\footnotesize,at={(0.5,0)},anchor=north,yshift=-1.55cm},
  legend style={font=\footnotesize,draw=none,legend columns=3}]
\nextgroupplot[title={(a) \SetValid selection criteria.},legend to name=confidencelegend]
  \addplot[black,solid,thick] table[x=threshold,y expr={1.25*\thisrow{A_precision}*\thisrow{A_recall}/(0.25*\thisrow{A_precision}+\thisrow{A_recall})},col sep=comma]{confidence_curves.csv};
  \addlegendentry{\(F_{0.5}\)}
  \addplot[black,dashed,thick] table[x=threshold,y=A_f1,col sep=comma]{confidence_curves.csv};
  \addlegendentry{\(F_1\)}
  \addplot[black,dash dot,thick] table[x=threshold,y expr={5*\thisrow{A_precision}*\thisrow{A_recall}/(4*\thisrow{A_precision}+\thisrow{A_recall})},col sep=comma]{confidence_curves.csv};
  \addlegendentry{\(F_2\)}
  \draw[gray,densely dashed] (axis cs:.95,0) -- (axis cs:.95,.7);
  \draw[gray,densely dashed] (axis cs:1.00,0) -- (axis cs:1.00,.7);
\nextgroupplot[title={(b) $F_1$ on each industrial set.},ylabel={},legend to name=setlegend]
  \addplot[black,solid,mark=*,thick] table[x=threshold,y=A_f1,col sep=comma]{confidence_curves.csv};
  \addlegendentry{\SetA}
  \addplot[black,dashed,mark=square*,thick] table[x=threshold,y=B_f1,col sep=comma]{confidence_curves.csv};
  \addlegendentry{\SetB}
  \addplot[black,dash dot,mark=triangle*,thick] table[x=threshold,y=C_f1,col sep=comma]{confidence_curves.csv};
  \addlegendentry{\SetC}
\end{groupplot}
\node at ($(group c1r1.south)+(0,-1.15cm)$) {\ref{confidencelegend}};
\node at ($(group c2r1.south)+(0,-1.15cm)$) {\ref{setlegend}};
\end{tikzpicture}
  \caption{Scores against the confidence threshold. Vertical dashed lines in (a) mark the selected thresholds.}\label{fig:confidence-curves}
\end{figure*}

Table~\ref{tab:confidence-thresholds} shows the unfiltered link set (No Filter) and the selected threshold rows from the full grid.
The Links column shows the number of predicted links retained at that threshold.
Figure~\ref{fig:confidence-curves} shows the validation criteria and the resulting \(F_1\) curves for all three industrial sets.

\begin{table}[t]\centering
  \caption{Selected Confidence Thresholds for \OurS on \SetValid}\label{tab:confidence-thresholds}
  \begin{tabular}{@{}lrrrrrrl@{}}\toprule
    $\tau$ & Links & $P$ & $R$ & $F_{0.5}$ & $F_1$ & $F_2$ & Selected for \\ \midrule
    No Filter & 313 & .288 & .486 & .313 & .361 & .427 & -- \\
    .95 & 210 & .419 & .476 & .429 & .446 & .463 & Max. $F_2$ \\
    1.00 & 117 & .667 & .422 & .597 & .517 & .455 & Max. $F_{0.5}$, $F_1$ \\ \bottomrule
  \end{tabular}
\end{table}

On \SetValid, \(F_{0.5}\) and \(F_1\) are both maximized at \(\tau=1.00\), while \(F_2\) is maximized at \(\tau=.95\).
At \(\tau=1.00\), filtering removes 196 false positives (67.1\%) and 18 true positives (26.9\%) from the combined unfiltered link set and raises \(F_1\) from .253 to .310 (\(1.23\times\)).
Across the combined curve for \SetTest, precision increases from .187 at \(\tau=0\) to .338 (\(1.81\times\)) at \(\tau=1\), while recall changes from .392 to .287 (\(0.73\times\)).
The control is coarse, however: the reported confidence values are concentrated at .90, .95, and 1.00, so the curves in Figure~\ref{fig:confidence-curves} are nearly flat below \(\tau=.90\), and the filter offers a few distinct link sets rather than a continuous trade-off.

The 31B model had no missing confidence tags in its 488 responses on \SetValid or its 504 responses on \SetTest with the selected \(B+M\) configuration.
In the model comparison (\RQ{4}), the 12B model also had no missing tags in 504 responses, while the E2B model had two in 504 responses.
Neither E2B response was parsed as a positive trace decision.
The missing-confidence fallback therefore contributed no true or false positives at \(\tau=1.00\) in these runs.

\RQAnswer{3}{Confidence provides a practical, if coarse, control for different tracing priorities. At \(\tau=1.00\), filtering removes \emph{67.1\% of false positives while retaining 73.1\% of true positives}, leaving fewer false-positive links for review. A lower threshold is available when missing links is more costly than reviewing extra candidates.}

\subsection{\RQ{4}: Model-Size Trade-Off}

\subsubsection{Motivation}
The preceding experiments with \ours use the primary 31B VLM, while tighter local hardware limits may require a smaller classifier.
To assess this deployment option, we need to determine how much recovery quality smaller checkpoints retain under the same pipeline settings.

\subsubsection{Design}
We run the Gemma 4 E2B, 12B, and 31B checkpoints separately on the full inputs of \SetB and \SetC.
Only the classifier checkpoint changes.
All three runs use the same multimodal embedding model, \(B+M\) candidate allocation, classifier input, prompt, parser, and decoding settings.
We also apply the same confidence threshold, \(\tau=1.00\), to every checkpoint.
This value maximizes \(F_1\) for the primary 31B VLM on \SetValid, and we do not tune a separate threshold for either smaller checkpoint.
For each checkpoint, we calculate precision, recall, and \(F_1\) on \SetB, \SetC, and their combination.
This design measures the performance change caused by substituting the classifier checkpoint under an otherwise fixed pipeline.

\subsubsection{Results and Discussion}
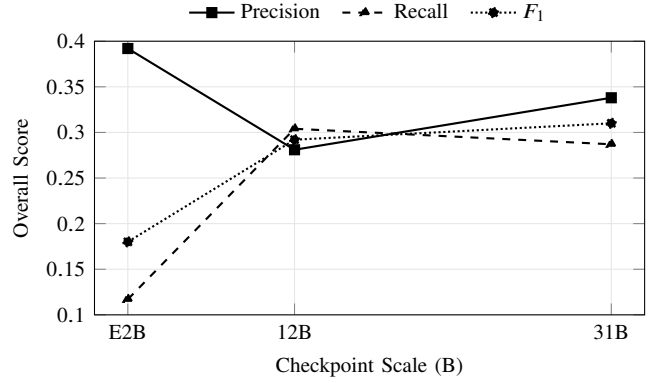
\begin{figure}[t]\centering
\begin{tikzpicture}
\begin{axis}[
  width=\columnwidth,
  height=5.2cm,
  xmin=0,xmax=33,
  ymin=.10,ymax=.40,
  xlabel={Checkpoint Scale (B)},
  ylabel={Overall Score},
  xtick={2,12,31},
  xticklabels={E2B,12B,31B},
  ytick={.10,.15,.20,.25,.30,.35,.40},
  grid=major,
  grid style={gray!20},
  tick label style={font=\footnotesize},
  label style={font=\footnotesize},
  legend style={
    at={(0.5,1.03)},
    anchor=south,
    draw=none,
    fill=none,
    font=\footnotesize,
    legend columns=3,
    /tikz/every even column/.append style={column sep=6pt}
  }]
  \addplot[black,thick,mark=square*,mark size=1.7pt] coordinates {(2,.392) (12,.281) (31,.338)};
  \addlegendentry{Precision}
  \addplot[black,thick,dashed,mark=triangle*,mark size=1.9pt] coordinates {(2,.117) (12,.304) (31,.287)};
  \addlegendentry{Recall}
  \addplot[black,thick,densely dotted,mark=*,mark size=1.7pt] coordinates {(2,.180) (12,.292) (31,.310)};
  \addlegendentry{$F_1$}
\end{axis}
\end{tikzpicture}
  \caption{Precision (squares), recall (triangles), and \(F_1\) (circles) for the three VLM sizes on the combined industrial test sets.}\label{fig:model-size}
\end{figure}

Figure~\ref{fig:model-size} compares precision, recall, and \(F_1\) across the three checkpoints.

The 31B checkpoint obtains combined \(F_1=.310\) (\(1.72\times\) E2B and \(1.06\times\) 12B) and is the highest-performing checkpoint on both \SetB and \SetC.
Compared with 12B, the 31B checkpoint recovers three fewer true positives but produces 37 fewer false positives, a reduction of 27.8\%.
The 12B checkpoint uses 38.7\% of the nominal parameter count of 31B while retaining 94.2\% of its combined \(F_1\).

\RQAnswer{4}{The 31B checkpoint gives the highest overall \(F_1\). The 12B checkpoint retains \emph{94.2\% of that \(F_1\) at 38.7\% of its nominal scale}, supporting use under tighter hardware limits. E2B retains 58.1\% of the 31B \(F_1\), but gives the highest precision. Larger size therefore improves overall \(F_1\), but not every tracing objective.}

\section{Discussion}\label{sec:discussion}

\subsection{Industrial Implications}
The results in Sections~\ref{sec:rq1} and~\ref{sec:rq2} show that image handling changes the task coverage of the executed approach: a text-only tool leaves the image-involving part of the task, 64.3\% of the test oracle, to manual checking, so image elements should be part of the input rather than reviewed separately.
The \(\tau=1.00\) link set of \ours recovers 21 image--image links and 16 of the 74 cross-modal gold links.
The final pipeline recovers links in both cross-modal directions.
The low image--text recall mainly comes from candidate retrieval: only 8 of the 60 image--text gold links enter the candidate sets, and the classifier retains 7 of them, giving a final recall of .117.
The classifier cannot recover gold links that are absent from the candidate set.
This pattern points to weak cross-modal semantic ranking for image queries in the shared embedding space.
Future work can compare global and within-modality ranks and similarity distributions, and examine task-specific embedding instructions, reranking, and embedding fine-tuning.

\subsection{Outputs for Different Tracing Priorities}
Thresholding provides smaller link sets with higher precision when a tracing task places more weight on review burden.
The \(F_{0.5}\), \(F_1\), and \(F_2\) criteria provide concrete selection points, while selecting a link set for a specific activity requires practitioner evaluation.

\subsection{Differences between Requirement Sets}
Image-involving gold links account for 34.4\% in \SetB and 100\% in \SetC (Table~\ref{tab:datasets}), showing different modality compositions in the two sets.
Both thresholds were selected on \SetA.
Lowering \(\tau\) from 1.00 to .95 reduces \(F_1\) on \SetB from .304 to .269 and increases it on \SetC from .320 to .333 (Table~\ref{tab:ours-by-set}).
Thus, the same threshold change has different effects on \(F_1\) across the test sets.
Other differences between the sets prevent us from isolating the effect of modality composition.

\subsection{VLM Selection}
Smaller VLMs remain useful when deployment hardware is limited.
The 12B checkpoint retains similar overall link quality to 31B, while E2B gives the highest precision despite its lower recall.
These results make deployment under tighter hardware limits more practical and provide a basis for studying lower-latency and interactive TLR.

\section{Threats to Validity}\label{sec:threats}

\subsection{Construct Validity}
Practitioners applied binary refinement or predecessor--successor relations to the studied HLR and LLR elements.
The confidence score is a self-reported model output used to derive binary link sets, not an oracle label or calibrated probability, and its values cluster at a few levels near 1.0.
Separate text--text and full-oracle evaluations distinguish input coverage from task-wide performance.
The semi-manual conversion preserves element order and represents each traceable unit as text or an image, but it may omit layout or other source-document structure.

\subsection{Internal Validity}
Each industrial requirement set was labeled by one practitioner, so no overlapping labels are available for measuring inter-annotator agreement.
Selecting the configuration and its \(F_1\) threshold on one validation set may favor \SetValid, although \SetTest is not used to revise these settings.
The selected allocation takes four candidates from each target modality and doubles the number per HLR from four to eight.
It may not suit other modality distributions, and its effect is not isolated from the larger candidate budget.
The comparison between the executed Hey et al. proxy and \ours changes the input modalities, embedding model, classifier, prompt, parser, and confidence output, making it an end-to-end comparison of the executed approaches.

\subsection{Conclusion Validity}
Each configuration is run once, and temperature zero does not guarantee identical GPU execution.
\RQ{4} compares selected checkpoints rather than isolating parameter count, and its threshold is selected for the primary VLM rather than each checkpoint.
The ranking may therefore change under another threshold.

\subsection{External Validity}
The industrial evidence uses three non-overlapping requirement sets from different parts of one system in one organizational setting.
The results may differ across organizations, languages, document structures, artifact granularities, image types, and model families.
The public evaluation uses six text-only req2req datasets from one package and configuration \(B\), which uses English instructions and no section metadata.

\subsection{Reproducibility}
The industrial datasets cannot be released publicly.
This limits exact replication of the industrial evaluation, while the public datasets and model checkpoints remain available.

\section{Conclusion}\label{sec:conclusion}

We studied whether a recent RAG-based inter-requirements TLR approach can recover links in industrial requirements containing both text and images.
\Ours recovers 49 of 171 industrial test links with precision .338, recall .287, and \(F_1=.310\).
This is 32 more (\(2.88\times\)) correctly recovered links than the executed text-only baseline.
The executed text-only baseline achieves an \(F_1\) score of 0.114, with lower precision and recall.
\Ours recovers links in all four modality combinations and reaches 33.6\% of image-involving gold links.
Text-only \(F_1\) improves on the industrial data and remains nearly unchanged on the public datasets.
Most gold links are still missed, and many predicted links are incorrect.
The recovered links should therefore be reviewed by practitioners before use.

Confidence filtering produces link sets with different precision and recall, allowing a choice according to tracing priorities.
Among the evaluated checkpoints, 31B gives the highest overall \(F_1\), 12B retains much of this performance, and E2B gives the highest precision.
Future work should improve retrieval of relevant text requirements for image queries, where many gold links are missed before classification.
Evaluation with practitioners and downstream tracing tasks is also needed to assess the usefulness of the recovered links.

\IEEEtriggeratref{15}

\end{document}